%% file: paper.tex
\documentclass{iopart}
\usepackage[left=2.5cm,right=2.5cm]{geometry}
\usepackage[T1]{fontenc}
\usepackage{ae}
\usepackage{aecompl}
\usepackage[square, comma, numbers, sort&compress]{natbib}
\usepackage{graphicx}
\usepackage{wrapfig}
\usepackage{color}
\usepackage{amssymb}
\usepackage{textcomp}
\usepackage{bm}

\begin{document}

\title{Combining Neutrino Oscillation Experiments with the Feldman-Cousins Method}

\author{A. V. Waldron$^{a}$, M. D. Haigh$^{a}$ and A. Weber$^{a,b}$}
\address{$^{a}$Oxford University, Department of Physics, Oxford, United Kingdom}
\address{$^{b}$STFC, Rutherford Appleton Laboratory, Harwell Oxford, United Kingdom}
\address{\texttt{a.waldron1@physics.ox.ac.uk}, \texttt{m.haigh1@physics.ox.ac.uk}, \texttt{alfons.weber@stfc.ac.uk}}

\begin{abstract}
In this article we describe how two or more experimental results can be combined within the procedure of Feldman and Cousins, to provide combined confidence limits on the physical parameters of interest.  We demonstrate the technique by combining the recent electron neutrino appearance results from T2K and MINOS.  Our best fit point is $\sin^{2}2\theta_{13} = 0.08 \,(0.11)$ and $\delta = 1.1 \,(2.0) \pi$; in addition we exclude $\sin^{2}2\theta_{13}=0$ at $2.7 \sigma$ ($2.8 \sigma$) for the normal (inverted) neutrino mass hierarchy.
\end{abstract}

\section{Introduction}
\input introduction.tex

\section{Method to Combine Fits}
\input method.tex

\section{Our Combination Example}

\input validation.tex

\input results.tex

\section{Conclusions}
\input conclusions.tex

\input bibliography.tex
\end{document}

%% file: introduction.tex
\noindent In order to obtain global constraints on physical parameters,
  it is often necessary to combine the results of two or more
  experiments with sensitivities to the same parameters.  Such
  combinations can be performed with different levels of
  sophistication, depending on what information is available about the
  original measurements. In the absence of detailed information,
  relatively crude combinations, \textit{e.g.}~by simply summing log-likelihood
  values, can be useful, but these results will always be subject to
  caveats.

  In this work, we describe a method to combine results which
  requires somewhat detailed experimental information, but which
  focuses on obtaining correct coverage of the parameter space under
  study, by producing a combined likelihood curve from the joint
  data set of the experiments, and using the Feldman-Cousins
  method~\citep{fc} to define acceptance regions for each point in the
  parameter space. We will describe the inputs that would be required
  from an experiment to enable its inclusion in such an analysis in
  \S~\ref{inputs}.  We also present, as an example of the technique,
  the combination of the recent electron neutrino appearance results
  from the long baseline experiments MINOS and T2K. This example is
  particularly pertinent, since the parameter $\theta_{13}$ is known
  to be near a physical bound; the Feldman-Cousins technique was
  developed to deal with such cases, since conventional methods for
  determining acceptance regions can produce incorrect coverage or
  null contours. A comparison will also be made between our method and
  one where fixed log-likelihood differences are used to produce the
  contours.

\subsection{The example: Constraints on $\theta_{13}$ from MINOS and T2K}

\noindent  Neutrino oscillations, a now well established
phenomenon~\citep{SNO,KamLAND,SK_atm,SK_tau,Minos_numu,Minos_numubar},
can be parametrised by the PMNS mixing matrix~\citep{P,MNS}, assuming a
minimal 3-flavour mixing model.  All of the mixing angles in this
matrix have now been measured to be non-zero, including the angle
$\theta_{13}$~\citep{dayaBay,reno}, which governs muon to electron neutrino oscillations in
the atmospheric ($L/E$) regime.  The $\mathcal{C\!P}$-violation
parameter $\delta$ is as yet unknown.  In the appearance channel T2K
has found an excess of electron neutrino events in a muon neutrino
beam, suggesting that $\theta_{13}$ is non-zero, with a significance
of $2.5\sigma$~\citep{t2kNue}.  MINOS has also found an excess of
events, excluding $\theta_{13}=0$ at 89\% CL~\citep{minosNue}.  We will
combine the results of these two experiments to get improved
constraints on $\theta_{13}$ from electron neutrino appearance
measurements, as well as the $\mathcal{C\!P}$-violation parameter
$\delta$.  In the disappearance channel, $\theta_{13}$ has recently been measured to be non-zero by two different reactor experiments, Daya Bay~\citep{dayaBay} and RENO~\citep{reno}.

MINOS and T2K are both long baseline accelerator neutrino oscillation
experiments, in which a predominantly $\nu_{\mu}$ beam is created by
colliding protons onto a target, then magnetically focusing the
resultant charged mesons and allowing them to decay. The neutrino beam
thus produced is sampled at source by a near detector, and several
hundred kilometers away by a far detector. Details of the individual
experiments' setups are described
elsewhere~\citep{minos_setup,t2k_setup}.  Both experiments are at an
atmospheric $L/E$, but their different energies and baselines (MINOS
is at 735 km and has a beam peaked at 3 GeV whereas T2K is at 295 km with
a beam peak of 0.6 GeV), along with different detector technologies and
analysis techniques lead to different systematic errors, as well as
different sensitivities to the oscillation parameters. Both
experiments use the
difference in the numbers of electron-neutrino-like events observed in
their far detectors from those predicted to constrain $\theta_{13}$.  Both analyses are performed with the Feldman-Cousins method.

%% file: method.tex
%% write method here...

\noindent

In order to combine the data from two or more experiments, the binned
data from all experiments are considered together, as a single larger
experiment. The Feldman-Cousins technique is then used to produce
confidence contours based on the combined data.

\subsection{Inputs to the analysis}
\label{inputs}
In order to perform our analysis, the data from each experiment are
required, along with Monte Carlo expectation values. Specifically,
these consist of:
\begin{enumerate} 

\item The number of data events in each analysis bin
for each experiment.

\item The expected number of events in each 
analysis bin, as a function of the parameters $\bm{\theta}$ to be
fitted over, in our example $(\sin^2 2\theta_{13},
\delta)$. The number of events will vary smoothly with
the oscillation parameters, so a set of values on a grid, which can be
numerically interpolated for intermediate points, is suitable for this
purpose.

\item The correlated bin-by-bin systematic errors for each experiment,
 encoded in covariance matrices, as a function of the parameters
 $\bm{\theta}$. We therefore require that the calculation of a
 covariance matrix from the underlying sources of systematic
 uncertainty (cross-section errors, beam-line uncertainties
 \textit{etc.}) has been performed by each experiment. We assume that
 both experimental and theoretical uncertainties are included in the
 matrix.

\end{enumerate}
Note that although we assume that the bin-to-bin correlations within
each experiment will be provided, in general there will also be
correlations between experiments, which may be important for the final
result. This detail is discussed in \S~\ref{corr_method}. It should be
noted further that in order to be combined consistently, we require
that the original analyses make essentially the same physics
assumptions; for example the values of mixing parameters other than
those in the fit.

\subsection{The Feldman-Cousins technique}
The Feldman-Cousins technique is a method for generating acceptance
regions within the classical framework for interval
calculation~\citep{neyman}. Its distinctive feature is that when
choosing which data values $\bm{n}$ to include in the acceptance for a
given $\bm{\theta}$, the candidates are ranked by the \emph{relative}
likelihood of observing $\bm{n}$ at $\bm{\theta}$, with respect to the
likelihood of observing $\bm{n}$ at the best fit point for that
data. The use of relative rather than absolute likelihoods means that
data values which are unlikely for all points in the parameter space
will still be included in the acceptance region for some
$\bm{\theta}$, avoiding the problem of null contours for some data
values. The method also automatically produces a smooth transition
between one- and two-sided intervals depending on the observed data,
with correct coverage throughout\footnote{Where the number of events
is very small, as with the T2K data, Feldman-Cousins will give some
over-coverage due to the discrete likelihood distribution generated
from toy Monte Carlo events. This effect will be negligible for the
combination example presented here.}.

The key procedural step in the technique is the generation of a large
number of ``toy'' Monte Carlo experiments at many positions in the
parameter space, in order to identify the relative likelihood limit
for each point which will give the correct coverage. For our example,
$10^4$ toy experiments were generated at each value of $\bm{\theta}$
considered. For each toy experiment result $\bm{n}$, a fit is
performed for $\bm{\theta}_\mathrm{best}$ using a likelihood function
$\ln
\mathcal{L}$. The difference, $\Delta (\ln
\mathcal{L})$, in the log likelihood between the best fit point
$\bm{\theta}_\mathrm{best}$, and the value $\bm{\theta}$ actually used
to generate the toy experiment, is calculated.

Using the $\Delta (\ln \mathcal{L})$ from all toy experiments, a value
$\Delta (\ln \mathcal{L})_\mathrm{crit}$ is calculated, such that some
fixed proportion, say 90\%, of toy experiments satisfy the condition
\begin{equation}
\Delta (\ln \mathcal{L}) \le \Delta (\ln
\mathcal{L})_{\mathrm{crit}}.
\label{eq::method::lnlcrit}
\end{equation}
The condition (\ref{eq::method::lnlcrit}) defines an acceptance region
with 90\% probability for the value of $\bm{\theta}$ in
question. Repeating the procedure for all values of $\bm{\theta}$, a
surface $\Delta (\ln
\mathcal{L})_\mathrm{crit}(\bm{\theta})$ is calculated.

A fit is then made to the real data using our likelihood function,
obtaining a best fit point $\bm{\theta}_\mathrm{best}$, and a
corresponding log-likelihood value $(\ln
\mathcal{L})_{\mathrm{best}}$. A log-likelihood surface $(\ln
\mathcal{L})(\bm{\theta})$ is also calculated on a grid in $\bm{\theta}$. 
We then draw our confidence interval, using the condition
(\ref{eq::method::lnlcrit}), with $\Delta (\ln
\mathcal{L})=(\ln
\mathcal{L})(\bm{\theta})-(\ln
\mathcal{L})_{\mathrm{best}}$, at each point in $\bm{\theta}$-space to decide whether
the point should be included in the contour.

\subsection{Toy experiment generation}
The toy Monte Carlo (MC) is required to give a number of events for
each analysis bin, $n_i^{\mathrm{obs}}$, based on the expected number
of events $n_i^{\mathrm{exp}}$ for a given value of $\bm{\theta}$,
allowing for fluctuations due to systematic~\footnote{Note that this method of treating systematics is Bayesian. In a frequentist prescription, systematics would be included as extra dimensions in $\bm{\theta}$; however this is not computationally feasible when using the Feldman-Cousins method.} and statistical
uncertainties. The expected number of events used is the total
(\textit{i.e.}~signal plus background) expectation, so that both
signal and background counts fluctuate between our toy experiments.
The main complication in the MC is to ensure that correlations between
the bin values are taken into account.

Expressing the systematic uncertainties as absolute shifts
(``tweaks'') $x_i$ in the expected number of events in each bin, we
can write
\begin{equation}
n_i^{\mathrm{exp}}\to n_i^{\mathrm{exp}}+x_i.
\end{equation}
Our method makes the common assumption that the systematic errors
follow a multivariate normal distribution; that is, their joint pdf
$f_{\mathrm{syst}}(\bm{x})$ follows
\begin{equation}
\label{eq::method::multigausspdf}
f_{\mathrm{syst}}(\bm{x})\propto e^{-\frac{1}{2}\bm{x}^{\top}V^{-1}\bm{x}},
\end{equation}
where $V$ is the covariance matrix for the $x_i$.

Since $V$ is a symmetric, positive-definite matrix, we can use the
method of Cholesky decomposition~\cite{chol} to find an upper-diagonal
matrix $L$ such that
\begin{equation}
\label{eq::method::choleskyvec}
V=L^{\top}L.
\end{equation}
We can then use the matrix $L^{\top}$ as a transformation to
enable us to generate the vector $\bm{x}$ from another vector of
random variables $\bm{y}$, such that
\begin{equation}
\bm{x}=L^{\top}\bm{y}.
\label{eq::method::tweakTransform}
\end{equation}
One can show by trivial matrix algebra that we can rewrite the factor
in the exponential in (\ref{eq::method::multigausspdf}) as
\begin{equation}
\bm{x}^{\top}V^{-1}\bm{x}=\bm{y}^{\top}\bm{y},
\end{equation}
and so we deduce that if we generate the $y_i$ independently on a
normal distribution with unit standard deviation, then the vector
$\bm{x}$ generated using (\ref{eq::method::tweakTransform}) will have
the desired covariance properties. From now on, unless otherwise
stated, we assume that $n_i^{\mathrm{exp}}$ has been modified for
systematics using this prescription, so that it is now a function
$n_i^{\mathrm{exp}}(\bm{\theta},\bm{y})$ of both $\bm{\theta}$ and
$\bm{y}$. We will also use $f_{\mathrm{syst}}(\bm{y})$ to refer to the
probability density function (pdf) of the random vector $\bm{y}$.

Having taken account of systematics, a number of events for each bin
can be generated using Poisson statistics, \textit{i.e.} on a pdf
\begin{equation}
g_{\mathrm{pois}}(n_i^{\mathrm{obs}};n_i^{\mathrm{exp}}(\bm{\theta},\bm{y}))=\frac{(n_i^{\mathrm{exp}}(\bm{\theta},\bm{y}))^{n_i^{\mathrm{obs}}}}{n_i^{\mathrm{obs}}!}e^{-n_i^{\mathrm{exp}}(\bm{\theta},\bm{y})},
\end{equation}
where the $y_i$ must be generated separately for each toy experiment.

\subsection{The likelihood function $\ln \mathcal{L}$}

When performing our fits, the value of the systematic ``tweak''
parameters is allowed to vary, so our likelihood function must include
penalty terms to account for the finite uncertainty in these
parameters. We define a likelihood function including both systematic
and statistical terms, as a function of $\bm{\theta}$ and $\bm{y}$,
using the same pdfs we used to generate the toy experiments:
\begin{eqnarray}
\mathcal{L}(\bm{n}^{\mathrm{obs}};\bm{y},\bm{\theta})&=
\prod_{i=0}^{N}\left(g_{\mathrm{pois}}(n_i^{\mathrm{obs}};n_i^{\mathrm{exp}}(\bm{\theta},\bm{y}))\right)f_{\mathrm{syst}}(\bm{y})\nonumber\\
&\propto
\prod_{i=0}^{N}\left((n_i^{\mathrm{exp}}(\bm{\theta},\bm{y}))^{n_i^{\mathrm{obs}}}e^{-n_i^{\mathrm{exp}}(\bm{\theta},\bm{y})}\right)
e^{-\frac{1}{2}\bm{y}^{\top}\bm{y}},
\end{eqnarray}
where we have dropped factors independent of
$(\bm{\theta},\bm{y})$. As is standard, minimisation is actually
performed on the logarithm of the likelihood function since this is a
simpler process, and is equivalent as $\ln \mathcal{L}$ is a monotonic
function of $\mathcal{L}$. This function is given by
\begin{equation}
-2\ln \mathcal{L} = \sum_{i=0}^{N}\left(2n_i^{\mathrm{exp}}(\bm{\theta},\bm{y})-2n_i^{\mathrm{obs}} \ln (n_i^{\mathrm{exp}}(\bm{\theta},\bm{y}))+y_i^2\right).
\label{eq::method::lnLDef}
\end{equation}
In the case that we do not wish to fit for systematics (\textit{e.g.} for
direct comparison with other analyses), we use the simpler function
\begin{equation}
-2\ln \mathcal{L} = 2\sum_{i=0}^{N}\left(n_i^{\mathrm{exp}}-n_i^{\mathrm{obs}} \ln n_i^{\mathrm{exp}}\right),
\end{equation}
where the $n_i^{\mathrm{exp}}$ are no longer functions of $\bm{y}$.

\subsection{Performing the fit}
We perform the minimisation of $(-2\ln \mathcal{L})$ using the
\texttt{MINUIT} minimiser via the interface provided by the
\texttt{ROOT} framework~\citep{root}. The derivatives $\frac{\partial \ln
 \mathcal{L}}{\partial y_i}$ are calculated analytically to improve
performance. The fit requires as inputs the vector
$\bm{n}^{\mathrm{exp}}$, and the Cholesky decomposition $L$ of its
covariance matrix $V$, at every point in $\bm{\theta}$ space. These
inputs are generated on a grid in $\bm{\theta}$ in advance and then
picked out by interpolation.

The fit is split into two nested components: a ``top-level fit''
over the parameters $\bm{\theta}$, and another fit performed over
$\bm{y}$ which is called by the top-level fit to find the best $(\ln
\mathcal{L})$ value for a given $\bm{\theta}$. $\sin^2(2\theta_{13})$
is constrained to lie in the physical region [0,1]. The $\bm{y}$ are
allowed to float freely, but all $n_i^{\mathrm{exp}}$ are constrained
to remain above zero (actually $10^{-5}$) regardless of the $\bm{y}$.

\subsection{Correlated systematic errors}
\label{corr_method}
When combining experiments with some common systematic error sources,
it is necessary to calculate the correlation coefficients between the
bins of the two experiments. In order to do this we must identify the
common systematic uncertainties, and get from each experiment the
contributions of these error sources to the total errors on each bin.

Once this information is available, the cross-terms in the covariance
matrix can be calculated. Assuming Gaussian errors, we express the
number of events in each bin $i$ as
\begin{equation}
n_i=\overline{n_i}+ \sigma^{\mathrm{uncorr}}_i y_i + \sum_{\alpha}c_{i \alpha}z_{\alpha},
\label{eq::method::corrSys}
\end{equation}
where the $y_i$ and $z_{\alpha}$ are independent normally distributed
random variables with ($\sigma=1$, $\mu=0$), $\alpha$ indexes the
correlated error sources and the $c_{i \alpha}$ give the contribution
of the error source $\alpha$ to the error on bin
$i$. $\sigma^{\mathrm{uncorr}}_i$ is the uncorrelated component of the
systematic error on bin $i$. From (\ref{eq::method::corrSys}), it is
easy to show that
\begin{equation}
V_{ij} \equiv \langle(n_i-\overline{n_i})(n_j-\overline{n_j}) \rangle
= \sum_{\alpha}c_{i \alpha}c_{j \alpha} +
\delta_{ij}(\sigma^{\mathrm{uncorr}}_i)^2.
\end{equation}

%% file: validation.tex
%% put the T2K validation results in here
\noindent In our example we combine the electron neutrino appearance results of
two long baseline experiments: MINOS~\citep{minosNue} and T2K~\citep{t2kNue}.  The inputs we take from
each experiment are the expected number of electron neutrino events
(as a function of $\theta_{13}, \delta$), in each bin, along with the
covariance matrices of the systematic errors between the bins in each
experiment. The T2K result consists of a single bin, and the MINOS
result of fifteen bins (five of energy times three of the Library
Event Matching particle identification parameter). 

\subsection{Validation}

To validate our method, we have reproduced the results of both the T2K
 and MINOS analyses individually. There are some slight differences in
 the ways in which the two experiments perform their analyses, which
 must be considered when validating our method, and when making a
 combination of their results.  One difference is that MINOS minimise
 over the systematic errors in their likelihood function, whereas T2K
 do not.  Since in general it may be advantageous to minimise over the
 systematic errors, we do minimise over systematics in our example
 fit.  Another difference is that T2K show their results for
 $\sin^{2}2\theta_{13}$ at a fixed value of $2\sin^{2}\theta_{23}=1$,
 whereas MINOS show the combination
 $2\sin^{2}\theta_{23}\sin^{2}2\theta_{13}$, by throwing $\theta_{23}$
 between its errors and then choosing $\theta_{13}$ to keep this
 quantity fixed in each toy experiment.  The systematic errors that
 MINOS have provided us with, however, do not include any uncertainty
 in the oscillation parameters (a fixed value of $2\sin^{2}\theta_{23}=1$ is used).  This means that our combination can
 be interpreted as constraining $\sin^{2}2\theta_{13}$ at
 $2\sin^{2}\theta_{23}=1$.  The other oscillation parameters used for calculation of the input distributions were: $\sin^{2}2\theta_{12} = 0.87$, $\Delta m^{2}_{21} = 7.6 \times 10^{-5}\, \textrm{eV}^{2}$ and $\Delta m^{2}_{32} = 2.3 \times 10^{-3}\, \textrm{eV}^{2}$.  For T2K, the provided inputs actually used $\Delta m^{2}_{32} = 2.4 \times 10^{-3}\, \textrm{eV}^{2}$, but changing to $\Delta m^{2}_{32} = 2.3 \times 10^{-3}\, \textrm{eV}^{2}$ made negligible difference to the result.
  
\begin{figure}
\begin{center}
\includegraphics[scale=0.7]{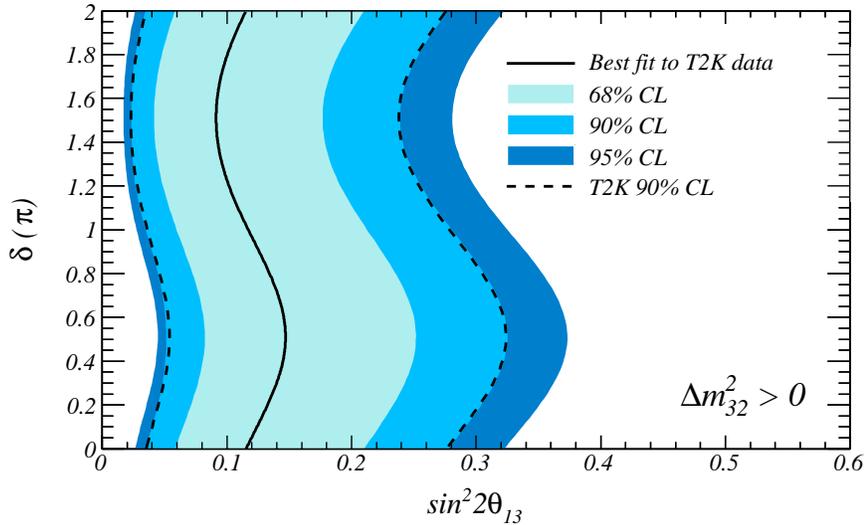}
\end{center}
\caption{Reproduction of the T2K results: we find good agreement with the published contours~\citep{t2kNue}. In addition, we exclude $\sin^{2}2\theta_{13}=0$ at $2.5\sigma$.}
\label{fig:t2kval}
\end{figure}

\begin{figure}
\begin{center}
\includegraphics[scale=0.7]{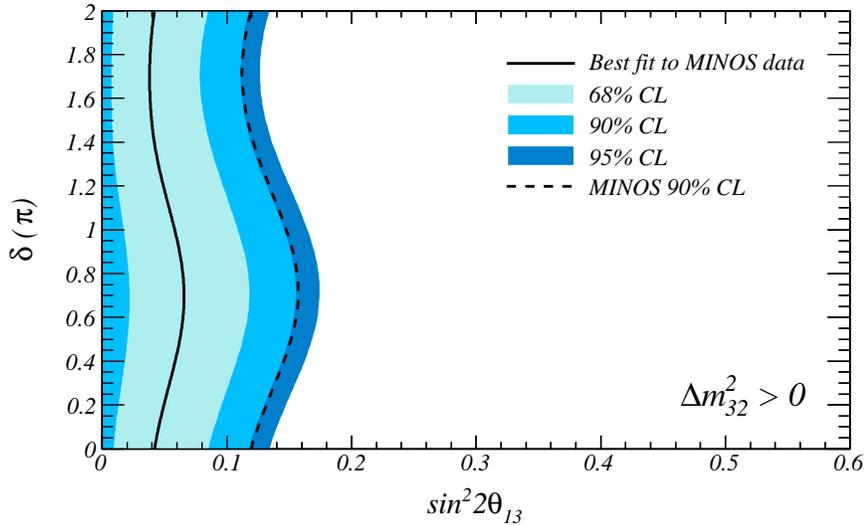}
\end{center}
\caption{Reproduction of the MINOS results: we find good agreement with the published contours~\citep{minosNue}.}

\label{fig:minosval}
\end{figure}

The validation results are shown in Figures~\ref{fig:t2kval} and
\ref{fig:minosval}, for T2K and MINOS respectively.  In both cases, we
find good agreement with the published contours.

%% file: results.tex
\subsection{Possibly Correlated Systematic Errors}
\label{corr}

In order to combine the errors from both experiments into a single
covariance matrix, we need to consider the systematic errors that may
be correlated between them. In the case of our example, most of the
sources of error cited in~\citep{minosNue,t2kNue} can be assumed to be
uncorrelated between experiments, due to the different experimental
set-ups and different neutrino energies involved.  However, cross
section uncertainties between the single T2K bin (energies up to 1.25
GeV) and the MINOS low energy (1-2 GeV) bins are potentially
correlated.

To assess the necessity of evaluating this correlation, we performed a
worst-case estimate of its effect, by assuming complete correlation
between the T2K cross-section error, and the total MINOS cross-section
plus flux error~\citep{minosOldNue}. These numbers were chosen as they
will overestimate the effect and are available from the cited
publications. Running our analysis with and without these worst-case
correlations included, we found negligible difference in the results.
We therefore conclude that we can neglect correlated errors for the
MINOS-T2K combination at this point, though this may not be true for
later data sets, if cross-section errors assume a greater relative
importance.

It should be noted for the future that whilst one reactor experiment
could be combined with MINOS and T2K without needing to account for
correlations (very different energies, flux, baseline
\textit{etc.}), if multiple reactor experiments were to be included
some effort would be needed to calculate the correlations between
their errors. The same would be true if another long-baseline neutrino
experiment with similar peak neutrino energy as either MINOS or T2K
were to be included.

\subsection{Results}

\begin{figure}
\begin{center}
\includegraphics[scale=0.7]{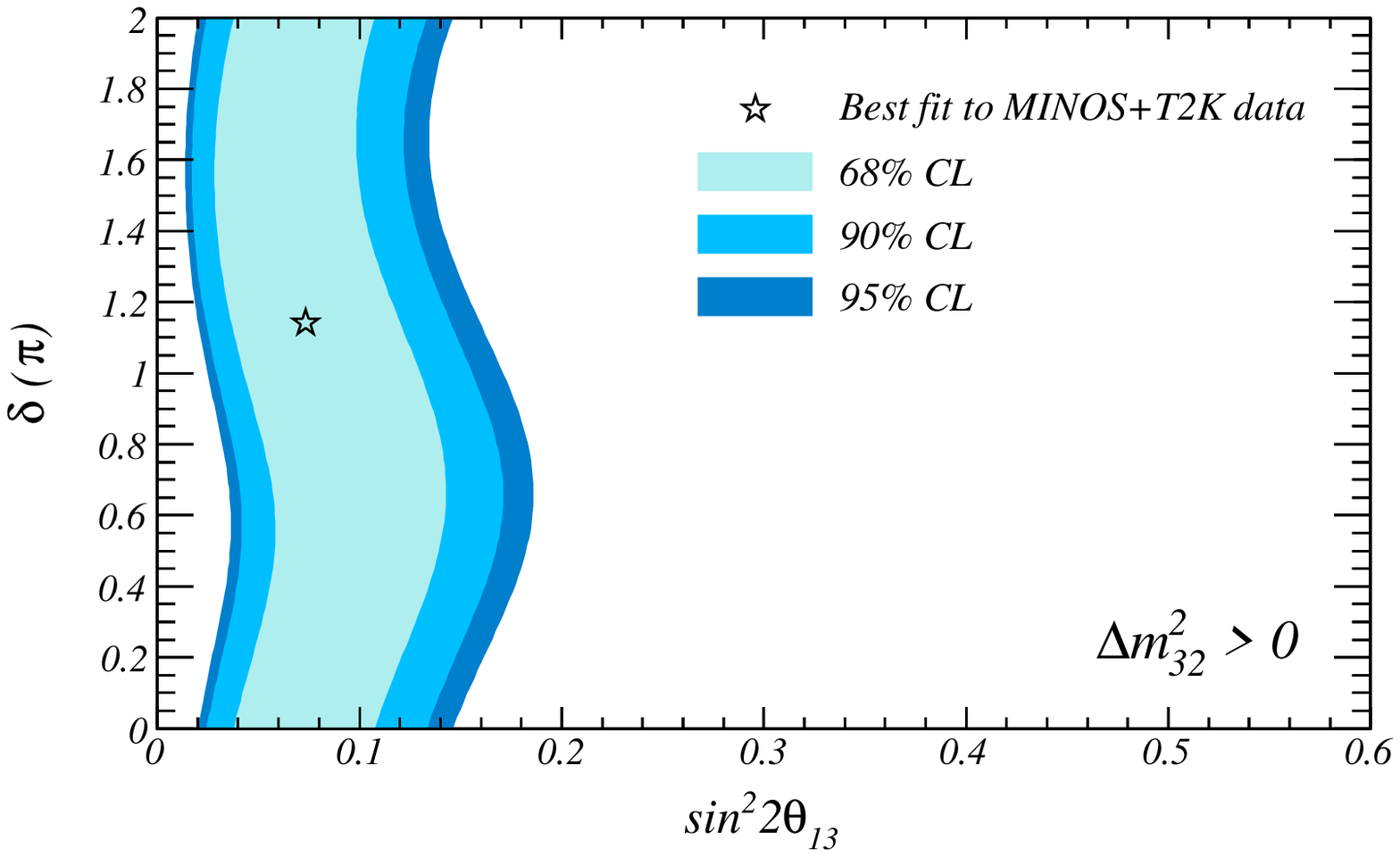}
\includegraphics[scale=0.7]{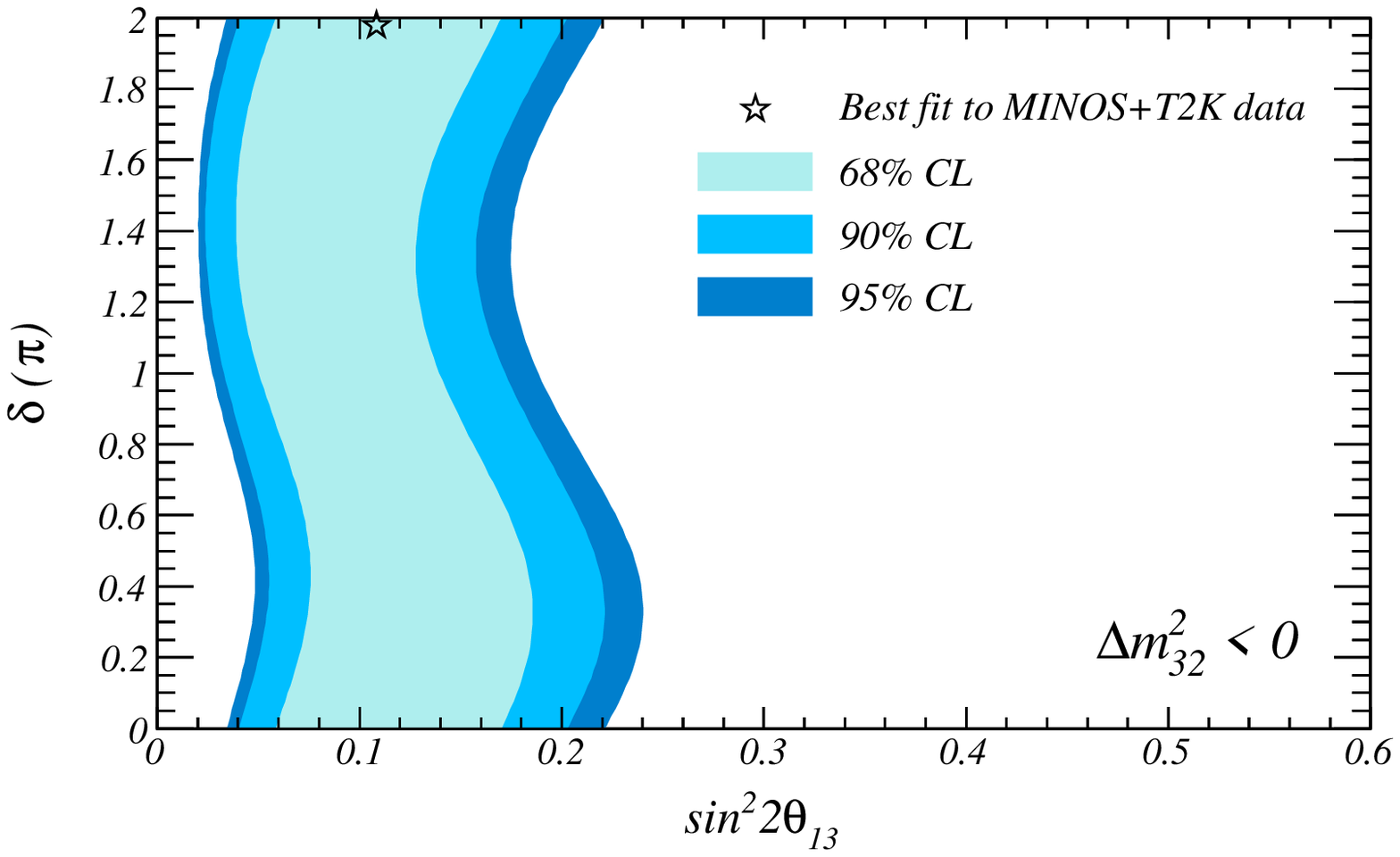}
\end{center}
\caption{The Feldman Cousins 68\%, 90\% and 95\% confidence level contours for the combined electron neutrino appearance measurement of T2K and MINOS, for normal (top) and inverted neutrino mass hierarchies.  We exclude $\sin^{2}2\theta_{13}=0$ at $2.7\sigma$, in the normal hierarchy.}
\label{fig:t2kdemo}
\end{figure}

\noindent As previously mentioned, we take the profile likelihood, minimising over systematic errors, in our final fit.  We also minimise over $\delta$ when calculating the minimum value of our likelihood function, since the combination of the two experiments gives slight sensitivity to $\mathcal{C\!P}$-violation.  The normal and inverted neutrino mass hierarchies are treated separately.  We find allowed regions of: $0.02 \,(0.03) < \sin^{2}2\theta_{13} < 0.16 \,(0.21)$ at 95\% C.L., $0.03 \,(0.04) < \sin^{2}2\theta_{13} < 0.15 \,(0.19)$ at 90\% C.L., and  $0.04 \,(0.05) < \sin^{2}2\theta_{13} < 0.12 \,(0.16)$ at 68\% C.L., for the normal (inverted) neutrino mass hierarchy, where we have taken the profile likelihood, minimising over $\delta$, for each value of $\sin^{2}2\theta_{13}$, both for the data and during the calculation of the critical values of $\Delta (2\textrm{ln} \mathcal{L})$ ($\delta$ was thrown uniformly across $[0,2\pi)$ in the generation of the toy experiments), to give one-dimensional confidence intervals.  Two-dimensional confidence intervals are shown in Figure~\ref{fig:t2kdemo}.  No values of $\delta$ can be ruled out at $1\sigma$.  The significance of the neutrino mass hierarchy preference, calculated by generating toy experiments about the global best fit point (which happens to be in the inverted neutrino mass hierarchy), and seeing what fraction of toy experiments had their global best fit point in the inverted mass hierarchy, is negligible.  The best fit values of the oscillation parameters are $\sin^{2}2\theta_{13} = 0.08 \,(0.11)$ and $\delta = 1.1 \,(2.0) \pi$ for the normal (inverted) neutrino mass hierarchy.  Our best fit value is compatible with the results presented in~\citep{schwetz}.  We exclude $\sin^{2}2\theta_{13}=0$ at $2.7 \sigma$ ($2.8 \sigma$) for the normal (inverted) neutrino mass hierarchy.

\begin{figure}
\begin{center}
\includegraphics[scale=0.7]{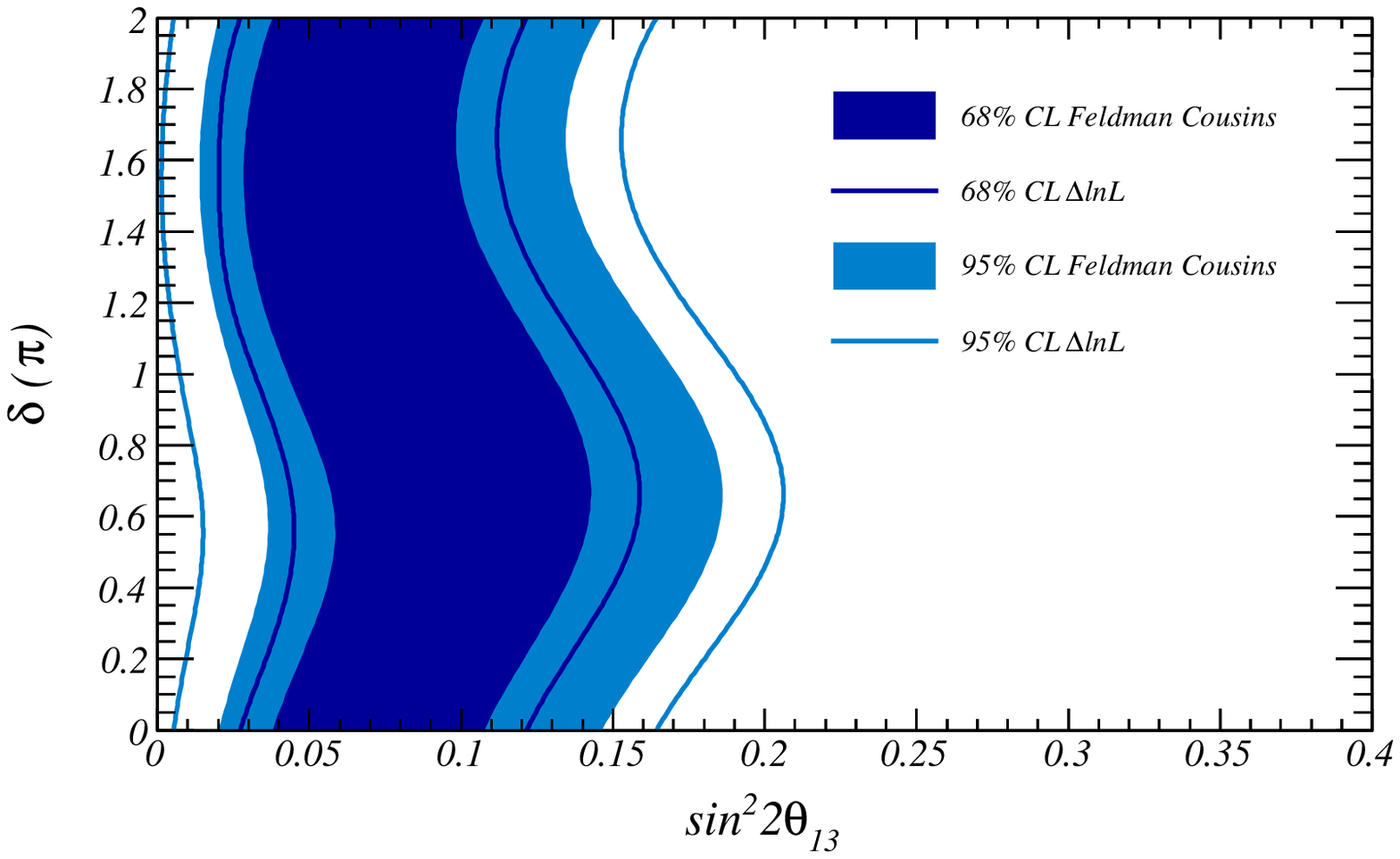}
\end{center}
\caption{The Feldman-Cousins confidence intervals and those obtained by placing the contour at a fixed $\Delta (2\textrm{ln} \mathcal{L})$ elevation.  Shown here are the 95\% CL allowed region (and corresponding $ \Delta (2\textrm{ln} \mathcal{L}) = 5.99$ contour) and 68\% CL allowed region (and corresponding $ \Delta (2\textrm{ln} \mathcal{L}) = 2.30$ contour).  The Feldman-Cousins allowed regions are significantly narrower than those formed using fixed $\Delta (2\textrm{ln} \mathcal{L})$ elevations.}
\label{fig:lnLcomp}
\end{figure}

For comparison, in Figure~\ref{fig:lnLcomp} we show the contours that
would be obtained by selecting an allowed region using a fixed value
of $\Delta (2\textrm{ln} \mathcal{L})$.  The regions obtained using
the Feldman-Cousins approach are significantly narrower than those from the
fixed log-likelihood contours.

%% file: conclusions.tex
\noindent We have demonstrated the combination of multiple experimental results with the Feldman-Cousins method.  Details of the inputs from experiments needed for inclusion in such fits are outlined in \S~\ref{inputs}.

We would like to thank the MINOS and T2K collaborations for their help with this work.  In particular we would like to thank: Ruth Toner and Lisa Whitehead, from MINOS, and Josh Albert, from T2K, for answering our many questions and helping with the validation of this work.  We are also indebted to Louis Lyons for useful discussions.  We acknowledge the support of STFC, U.K.